\pdfoutput=1

\documentclass[11pt]{article}

\usepackage[preprint]{coling}

\usepackage{times}
\usepackage{latexsym}

\usepackage[T1]{fontenc}

\usepackage[utf8]{inputenc}

\usepackage{microtype}

\usepackage{inconsolata}

\usepackage{graphicx}

\title{GenCRF: Generative Clustering and Reformulation Framework for Enhanced Intent-Driven Information Retrieval}

\author{
 \textbf{Wonduk Seo}\textsuperscript{1,2}\thanks{equal contribution.}\hspace{0.5em}
 \textbf{Haojie Zhang}\textsuperscript{1}$^{*}$\hspace{0.5em} 
 \textbf{Yueyang Zhang}\textsuperscript{1}$^{*}$\hspace{0.5em}
 \textbf{Changhao Zhang}\textsuperscript{1,2}\\
 \textbf{Songyao Duan}\textsuperscript{1,2}\hspace{0.5em}
 \textbf{Lixin Su}\textsuperscript{1}\hspace{0.5em}
 \textbf{Daiting Shi}\textsuperscript{1}\hspace{0.5em}
 \textbf{Jiashu Zhao}\textsuperscript{3}\hspace{0.5em}
 \textbf{Dawei Yin}\textsuperscript{1}\thanks{corresponding author.}
\\
\\
 Baidu.inc\textsuperscript{1}\hspace{0.5em}
 Peking University\textsuperscript{2}\hspace{0.5em}
 Wilfrid Laurier University\textsuperscript{3}\\ 
 \{seowonduk\}@pku.edu.cn, \{2301210522, duansy\}@stu.pku.edu.cn\\
 \{zhanghaojie03, zhangyueyang, sulixin, shidaiting01, yindawei02\}@baidu.com, 
 \{jzhao\}@wlu.ca
}
 
\usepackage{algorithm}
\usepackage{amsmath}
\usepackage{algpseudocode}
\usepackage{pifont}
\usepackage{booktabs}
\usepackage{multicol}
\usepackage{array}
\usepackage{multirow}
\usepackage[export]{adjustbox}
\usepackage{pdflscape}

\begin{document} 
\maketitle
\begin{abstract}
Query reformulation is a well-known problem in Information Retrieval (IR) aimed at enhancing single search successful completion rate by automatically modifying user's input query. Recent methods leverage Large Language Models (LLMs) to improve query reformulation, but often generate insufficient and redundant expansions, potentially constraining their effectiveness in capturing diverse intents. In this paper, we propose \emph{\textbf{GenCRF: a Generative Clustering and Reformulation Framework}} to capture diverse intentions adaptively based on multiple differentiated, well-generated queries in the retrieval phase for the first time. GenCRF leverages LLMs to generate variable queries from the initial query using customized prompts, then clusters them into groups to  distinctly represent diverse intents. Furthermore, the framework explores to combine diverse intents query with innovative weighted aggregation strategies to optimize retrieval performance and crucially integrates a novel Query Evaluation Rewarding Model (QERM)  to refine the process through feedback loops. Empirical experiments on the BEIR benchmark demonstrate that GenCRF achieves state-of-the-art performance, surpassing previous query reformulation SOTAs by up to $12$\% on nDCG@10. These techniques can be adapted to various LLMs, significantly boosting retriever performance and advancing the field of Information Retrieval.

\end{abstract}

\section{Introduction}

Query reformulation is a well-known problem in Information Retrieval (IR) to enhance search effectiveness by automatically modifying the initial query into well-formed one(s)\cite{carpineto2012}. Traditional Pseudo-Relevance Feedback (PRF) based methods, such as RM3, improve the initial query by selecting terms from relevant documents \cite{Robertson1991,Lavrenko2001}. Similarly, researchers expand initial queries by incorporating semantically similar terms with pre-trained word embeddings \cite{kuzi2016,roy2016,zamani2016}. With the advent of Large Language Models (LLMs), query reformulation has re-emerged as a prominent research area within the field of information retrieval \cite{zhao2023a}. In contrast to past methods that relied on using existing related terms in the retrieval system for expansion, the current approaches to query reformulation harness the exceptional generative understanding abilities of LLMs \cite{wang2023a, li2023}. They leverage foundational LLM techniques such as prompt engineering and Chain-of-Thought (CoT) to enhance initial queries by generating keywords and detailed descriptions \cite{wei2022, jagerman2023}. However, these methods often face limitations in enriching information capacity through single expansions.

More recently, ensemble approaches utilizing multiple prompts to generate various keywords have emerged, demonstrating improved performance compared to earlier single expansion methods \cite{li2023, dhole2024a, dhole2024b}.  Although these methods demonstrate the benefits of utilizing various expansions to enrich original queries and improve retrieval effectiveness, these methods face several challenges: \ding{172} The variations in their prompts tend to be simplistic and homogeneous prompt variations, lacking effective methods to capture the diverse user intents from multiple perspectives, \ding{173} These methods primarily lack of dynamic assessment of intent importance and query relevance, \ding{174} There is a lack of effective mechanisms to detect generation quality, potentially introducing negative biases in query performance.

To overcome these limitations, we propose GenCRF: a Generative Clustering and Reformulation Framework. Unlike previous methods that generate keywords or documents, GenCRF directly leverages LLMs to generate multiple differentiated queries derived from the original input by utilizing various types of customized prompts. Through detailed analysis and observation, we identified several query expansion types and designed customized prompts: "contextual expansion," "detail specific," and "aspect specific". GenCRF then dynamically clusters these queries to capture diverse intents, minimizing information redundancy and maximizing the potential of query reformulation.

In order to efficiently integrate abundant and diversified multi-intent queries, GenCRF incorporates several weighted aggregation strategies, including similarity-based dynamic weighting \emph{(GenCRF/SimDW)} and score-based dynamic weighting \emph{(GenCRF/ScoreDW)}, to adjust the relative weights of reformulated queries based on various criteria and efficiently integrate diverse multi-intent queries. To further enhance performance, we introduce a fine-tuning step \emph{(GenCRF/ScoreDW-FT)} that optimizes the model's ability to evaluate and score reformulated queries. Ultimately, we introduce the Query Evaluation Rewarding Model \emph{(QERM)}, which evaluates clustering generation quality and guides query refinement through a feedback loop. QERM subsequently guides the LLMs to either continue refining the queries or conclude the process as appropriate, ensuring optimal refinement and high-quality query formulation.

Extensive experiments on the BEIR dataset \cite{thakur2021beir} through competitive LLMs demonstrate GenCRF 's consistent superiority over state-of-the-art query reformulation techniques across diverse domains and query types. Comprehensive analyses of initial query weight, prompt quantity, number of generated queries and QERM iteration count further validate GenCRF's effectiveness. Our investigations confirm GenCRF's robustness, capacity to generate highly diverse results, and ability to effectively cluster and retrieve a wide spectrum of intents. These findings not only validate our approach but also offer valuable insights for future information retrieval research.

\begin{figure*}[!htpb]
    \centering
    \includegraphics[width=1\textwidth]{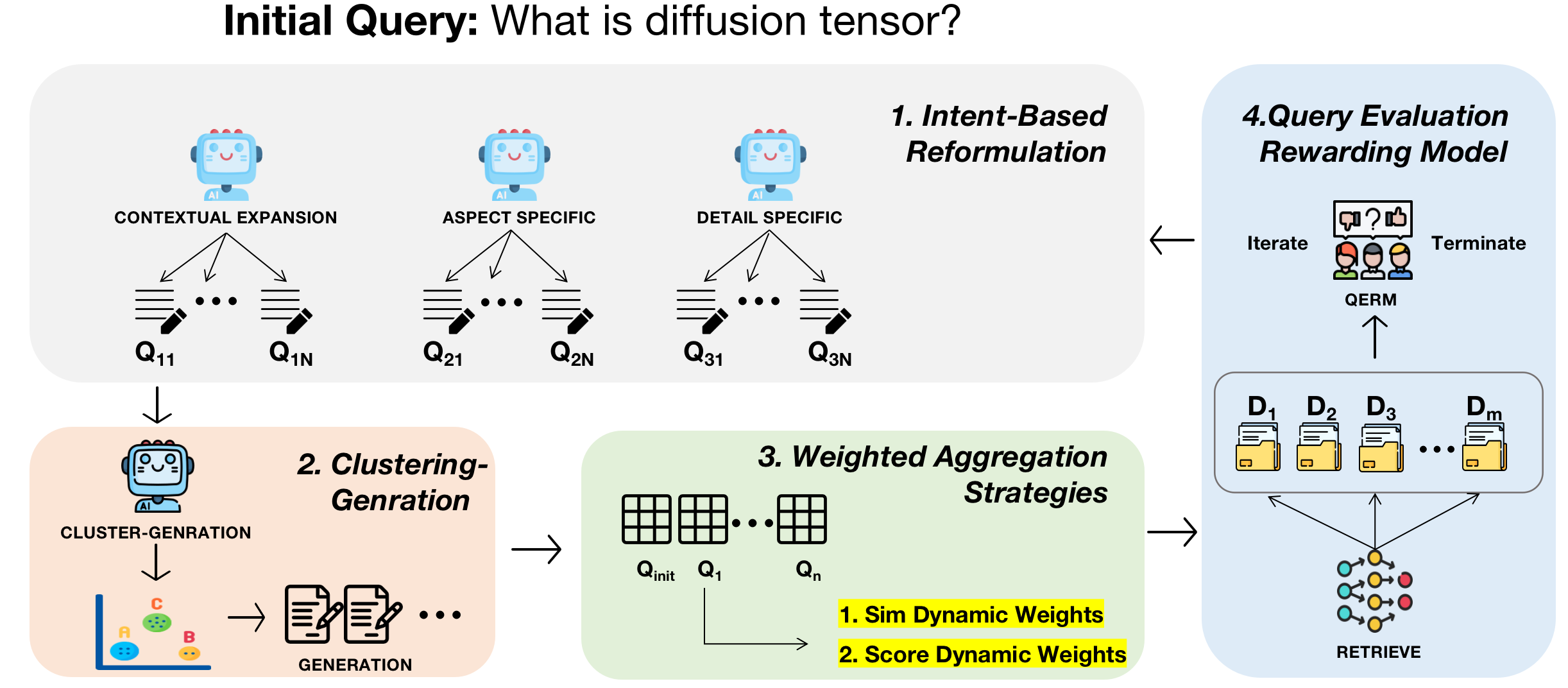}
    \caption{Overview of the GenCRF: Generative Clustering and Reformulation Framework}
    \label{fig:GenCRF_Overview}
\end{figure*}
\section{RELATED WORK}
Numerous methods have been applied for query reformulation, which has significantly evolved over the years, adapting to new methodologies in information retrieval (IR). Early approaches relied on classical retrieval models such as BM25 \cite{robertson1994}, which focused on exact matching statistical features including term frequency and document length to assess relevance. These methods often utilize techniques such as RM3 and query logs for pseudo-relevance feedback \cite{Robertson1991,Lavrenko2001, jones2006, craswell2007}. Neural networks have provided a new perspective on developing more sophisticated methods for query reformulation. Grbovic et al. (\citeyear{grbovic2015}) proposed a rewriting method based on a query embedding algorithm, and Nogueira et al. (\citeyear{nogueira2017}) also explored reinforcement learning-based models. Dense neural networks further advanced the field of query reformulation, with pre-trained embeddings, capturing complex semantics and facilitating transfer learning in IR tasks \cite{devlin2019, xiong2021}.

More recently, Large Language Models (LLMs) have significantly transformed query reformulation strategies. Weller et al. (\citeyear{weller2024}) emphasized the potential of LLMs to utilize their ability including query reformulation, and showed that LLMs outperform traditional methods for query expansion. Generating keywords and pseudo documents such as Query2Doc (Q2D) \cite{wang2023a} and Query2Expansion (Q2E) \cite{jagerman2023} have shown their effectiveness in improving retrieval quality \cite{nogueira2019, claveau2022, wang2023b}. Although those methods have shown promise in query reformulation to some extent, they often rely on a single model or prompt. To address the limitations of previous query reformulation methods, recent studies found that applying multiple different prompts to generate various keywords or documents could further boost the overall quality of query reformulation, as they can provide a certain degree of information gain for retrieval queries more closely, thereby effectively capturing a broader range of user intent. \cite{ li2023, dhole2024a}. 

Despite their improvements, above methods still face formidable challenges. They often tend to employ simplistic prompt variations that may not adequately capture the breadth of diverse user intents, resulting in redundant keyword generations that undermine the effectiveness of query reformulation. Moreover, their ensemble techniques frequently fall short in appropriately emphasizing the significance of various intents and in dynamically weighting the relevance between initial queries and reformulated ones. 
There is also a noticeable absence of robust mechanisms to evaluate the quality of the generated outputs, which can result in the inclusion of semantically ambiguous terms, ultimately detracting from the overall performance.

\section{METHODOLOGY}
In this section, we first provide a comprehensive overview of our innovative  \emph{\textbf{Generative Clustering and Reformulation Framework (GenCRF)}} (Section 3.1), followed by our specific Generation \& Clustering settings and a comparative analysis with existing methods (Section 3.2). We then present weighted aggregation and fine-tuning strategies to optimize retrieval performance (Section 3.3). Finally, we introduce our Query Evaluation Rewarding Model (QERM), desgined to further enhance GenCRF's performance through intent-driven query capture and critical feedback for query re-generation and re-clustering (Section 3.4). 

\subsection{Overview of GenCRF} \label{subsec:Overview of GenCRF}
To construct the GenCRF, we first utilize LLMs to reformulate the initial query $q_{\text{init}}$, into a new form $q_{\text{new}}$. This process generates N queries for each of the 3 diverse customized prompts in set $P$:
\begin{equation}
\mathcal{Q}_{\text{gen}} = \bigcup_{\text{prompt} \in P} \{R(q_{\text{init}}, \text{prompt})\}
\end{equation}
In this equation, $\mathcal{Q}_{\text{gen}}$ represents the set of all generated queries. The reformulation LLM, denoted as $R$, applies each prompt in $P$ to the initial query. To reduce information redundancy and capture diverse intents, we introduce a clustering step in our framework. This step dynamically clusters generated queries into several intentional groups and produces a representative, comprehensive query for each cluster:
\begin{equation}
\mathcal{Q}_{\text{final}} = \text{G}(q_{\text{init}}, \mathcal{Q}_{\text{gen}})
\end{equation}
The function $\text{G}$ clusters the set of generated queries $\mathcal{Q}_{\text{gen}}$ into 1 to 3 groups dynamically. It then generates a new representative query for each cluster, resulting in the set $\mathcal{Q}_{\text{final}}$. This procedure ensures comprehensive coverage of derived query intents beyond the original query, while eliminating similar or redundant queries. Following the clustering step, the framework proceeds with a retrieval process. This process combines various weighted aggregation strategies designed to effectively capture both $q_{\text{init}}$ and $\mathcal{Q}_{\text{final}}$:
\begin{equation}
\mathcal{D}_{\text{retrieval}} = \text{Retrieve}(\mathcal{Q}_{\text{final}}, q_{\text{init}}, \mathcal{W})
\end{equation}
where $\mathcal{W}$ represents the weighting parameters used in the aggregation strategies and $\mathcal{D}_{\text{retrieval}}$ is the set of final retrieved documents. To further enhance its performance, the GenCRF framework incorporates a novel Query Expansion Rewarding Model (QERM), which detects the superiority of clustered intent-driven queries and provides effective feedback to LLMs, signaling when re-generation and re-clustering are necessary. The overall pipeline of the GenCRF is shown in Figure \ref{fig:GenCRF_Overview}.

\subsection{Generation and Clustering}
\label{section:3.2}
Prompts used by current baselines, such as Query2Doc (Q2D) and Query2Expansion (Q2E), typically instruct models to produce relevant keywords or documents without considering the inherent intent and underlying value of the query. Moreover, these prompts often exhibit simplicity and homogeneity, so that lack of the depth required to effectively capture diverse user intents. 

Through comprehensive analysis and observation, we have identified several distinct query expansion intents: \emph{Contextual Enrichment} broadens queries with relevant context; \emph{Detail-Oriented Exploration} focuses on specific subtopics; \emph{Aspect-Focused Expansion} concentrates on particular facets; \emph{Clarification-Focused Refinement} clarify ambiguities; and \emph{Exploratory Intent} investigates related but unexplored areas. From these observations, we devise three types of tailored and effective intents to diversify generated queries from multiple perspectives as follows:
\vspace{-2pt} 
\begin{description}
    \setlength{\itemsep}{0pt}
    \setlength{\parskip}{0pt}
    \item[\emph{1. Contextual Expansion:}]
    Expands the initial query's context while maintaining clarity, ensuring comprehensive understanding and generating more relevant, refined reformulations.
    \item[\emph{2. Detail Specific:}]
    Elicits specific details or subtopics within the query, providing focused insights and enhancing the granularity of retrieved information.
    \item[\emph{3. Aspect Specific:}]
    Concentrates on a specific aspect or dimension of the topic, broadening the query's scope while focusing on the target dimension to enrich result diversity.
\end{description}
\vspace{-2pt} 
 
To further enhance query diversity while maintaining focus on core intent, we propose a clustering generation prompt to guide the LLM to explore multi-type demands, as follows:
\begin{description}
    \setlength{\itemsep}{0pt}
    \setlength{\parskip}{0pt}
    \item[\emph{4. Clustering-Generation:}]
    Extracts up to three intent queries from differentiated queries in GenCRF, enriching the query reformulation process, improving overall query intent understanding and reformulation strategies. 
\end{description}
\subsection{Weighted Aggregation Strategies}
\label{section:Weighted Aggregation Strategies}
In order to optimize retrieval performance by effectively capturing both $q_{\text{init}}$ and reformulated queries from $\mathcal{Q}_{\text{final}}$, we introduce two distinct weighted aggregation strategies and a fine-tuning process.

\paragraph{\textbf{Similarity Dynamic Weights (SimDW).}}
This novel strategy dynamically adjusts the weights of reformulated queries based on their similarity to $q_{\text{init}}$, while incorporating a filtering mechanism to ensure relevance. After assigning a fixed weight to the initial query, the method considers only those reformulated queries exceeding a predefined similarity threshold in the dynamic weighted aggregation. The aggregation equation is given by:
\vspace{-2pt} 
\begin{equation}
q_{\text{agg}}^{\text{simDW}} = w_0 \cdot q_{\text{init}} + \sum_{\substack{i=1 \\ \text{sim} \geq \theta}}^{|Q_f|} \text{sim}(q_{\text{init}}, q_{f,i}) \cdot q_{f,i}
\end{equation}
where $w_0$ is the fixed weight for the initial query; $\text{sim}$ represents a dynamic weight estimating the relative magnitude of $q_{\text{f},i}$, calculated as the cosine similarity between the embeddings of the initial query and the $i$-th reformulated query $q_{\text{f}, i}$ using a sentence embedding model; and $\theta$ is the similarity threshold for filtering irrelevant queries.

\paragraph{\textbf{Score Dynamic Weights (ScoreDW).}}
Building upon the SimDW approach, the ScoreDW strategy offers a more comprehensive evaluation of reformulated queries by employing a multidimensional scoring system to assess query quality, using these scores as dynamic weights in the aggregation process. The method retains the fixed weight for the initial query and the filtering mechanism from SimDW, but enhances the evaluation criteria. The aggregation equation for ScoreDW is expressed as:
\vspace{-2pt} 
\begin{equation}
q_{\text{agg}}^{\text{scoreDW}} = w_0 \cdot q_{\text{init}} + \sum_{\substack{i=1 \\ \text{score} \geq \theta}}^{|Q_f|} \text{score}(q_{\text{init}}, q_{f,i}) \cdot q_{f,i}
\end{equation}

Specifically, $\text{score}$ is a dynamic weight representing the estimated importance of each $q_{\text{f},i}$, derived from an LLM's evaluation of the reformulated query relative to the initial query. The evaluation considers five key dimensions: Relevance, Specificity, Clarity, Comprehensiveness, and Usefulness for retrieval. The threshold $\theta$ ensures that only high-scoring, pertinent reformulations contribute to the final aggregated query.

\paragraph{\textbf{Fine-Tuning for ScoreDW.}} 
For the purpose of optimizing the ScoreDW strategy, we implement a fine-tuning process for the LLMs to enhance their precision in evaluating and scoring reformulated queries. The process begins with the generation of a diverse set of query pairs $(q_{\text{init}}, q_{\text{ref}})$ using each LLM, where $q_{\text{ref}}$ is the reformulated query. These pairs are then evaluated by GPT-4o, serving as a high-quality benchmark, to produce reference scores. The fine-tuning objective is formulated as:
\begin{equation}
\phi^* = \arg \min_{\phi} \sum_{i=1}^{N} \mathcal{L}(\text{LLM}_{\phi}(q_{\text{init},i}, q_{\text{ref},i}), s_i)
\end{equation}
where $\phi^*$ denotes the optimal LLM parameters, $(q_{\text{init},i}, q_{\text{ref},i})$ represents the $i$-th query pair, $s_i$ is the corresponding score generated by GPT-4o, and $\mathcal{L}$ is the loss function. Fine-Tuning process aims to enhance the LLM's ability to discriminate between high and low-quality reformulations, ensuring consistent and scalable query quality assessment.

\subsection{Query Evaluation Rewarding Model}
\label{section:Query Evaluation Rewarding Model}

To further improve the performance of GenCRF, we also introduce a novel approach: the Query Evaluation Rewarding Model (QERM). This innovative model functions as a multi-intent gain detection model that assesses the quality and effectiveness of queries generated by GenCRF, focusing on their alignment with diverse, intent-driven clusters. QERM evaluates how well generated queries capture user intent and contribute to meaningful query clusters. It provides feedback to LLMs for re-generation and re-clustering if necessary, addresses limitations in initial scoring.
\begin{algorithm}
\begin{small}
\caption{Query Evaluation Rewarding Model}\label{algo2}
\begin{algorithmic}[1]
\Require  nDCG threshold $\tau$, output logit threshold $\varepsilon$, training dataset with Queries $Q = \{q_1, q_2, \ldots, q_n\}$, Maximum Iteration $M$
\State // Construct training dataset for reward model
\For{each $q \in Q$}
    \State Implement Generation, Clustering and Weighted Aggregation in the GenCRF for $q$
    \State Compute $\text{nDCG@10}(q)$ from retrieval documents
    \If{$\text{nDCG@10}(q) < \tau$}
        \State \Return $\text{label}(q) \Leftarrow 0$
    \Else
        \State \Return $\text{label}(q) \Leftarrow 1$
    \EndIf
\EndFor
\State // Training
\State Train Reward Model with labeled datasets to assess the superiority of clustered intent-driven queries
\State // Inferring
\State Initialize timestep $t \Leftarrow 0$
\While{$t < M$}
    \State Provide feedback from the output logit produced by reward model
    \If{$\text{output logit} < \varepsilon$}
        \State Implement re-Generation and re-Clustering in the GenCRF
    \Else
        \State \Return retrieval results
    \EndIf
    \State $t \Leftarrow t+1$
\EndWhile
\end{algorithmic}
\end{small}
\end{algorithm}

QERM calculates nDCG@10 scores for each query, assigning labels based on a threshold ($\varepsilon$). Queries below the threshold are labeled as "0" for re-generation, while those above are labeled as "1", denoting satisfactory performance expected. A language model is then trained on these labeled queries to guide query refinement decisions. The trained reward model is subsequently used to infer the query quality in the test set, providing critical feedback for re-generation and re-clustering as described in Algorithm \ref{algo2}, thereby ensuring high query quality standards and improving retrieval performance.

\begin{table*}[t]
\makebox[\textwidth][c]{%
\resizebox{1\textwidth}{!}{%
\begin{tabular}{@{}cclccccccc@{}}
\toprule[1pt]
\textbf{Model} & \textbf{Approach} & \textbf{Methods} & \textbf{scifact} & \textbf{trec-covid} & \textbf{scidocs} & \textbf{nfcorpus} & \textbf{dbpedia-entity} & \textbf{fiqa} & \textbf{Avg.} \\
\midrule 
\multirow{9}{*}{\centering{\textbf{Mistral}}} & \multirow{7}{*}{\centering{non-fine-tuned}} & Q2D & 0.5966 & 0.5669 & 0.1316 & 0.2964 & 0.3750 & 0.2645 & 0.3718 \\
& & Q2E & 0.5895 & 0.5941 & 0.1267 & 0.2863 & 0.3200 & 0.2677 & 0.3641 \\
& & CoT & 0.5986 & 0.5978 & 0.1314 & 0.2916 & 0.3740 & 0.2418 & 0.3725 \\
& & GenQRE & 0.6306 & 0.6062 & 0.1490 & 0.2844 & 0.3688 & 0.2353 & 0.3791 \\
& & GenQRFusion & 0.6370 & 0.5474 & 0.1406 & 0.3068 & 0.3662 & 0.2701 & 0.3780 \\
& & GenCRF/SimDW\textsuperscript{$\dagger$} & \underline{0.6629}\textsuperscript{$*$} & 0.6447\textsuperscript{$*$} & \underline{0.1529}\textsuperscript{$*$} & 0.3268 & 0.3867\textsuperscript{$*$} & \underline{0.3083}\textsuperscript{$*$} & 0.4137\textsuperscript{$*$} \\
& & GenCRF/ScoreDW\textsuperscript{$\dagger$} & 0.6574\textsuperscript{$*$} & 0.6451\textsuperscript{$*$} & 0.1521 & 0.3267\textsuperscript{$*$} & 0.3862 & 0.3068\textsuperscript{$*$} & 0.4124\textsuperscript{$*$} \\
\cmidrule{2-10}
& \multirow{2}{*}{\centering{fine-tuned}} & GenCRF/ScoreDW-FT\textsuperscript{$\dagger$} & 0.6596\textsuperscript{$*$} & \underline{0.6527}\textsuperscript{$*$} & 0.1523 & \underline{0.3270} & \underline{0.3873}\textsuperscript{$*$} & 0.3077\textsuperscript{$*$} & \underline{0.4143}\textsuperscript{$*$} \\
& & GenCRF/ScoreDW-FT-QREM\textsuperscript{$\dagger$} & \textbf{0.6637}\textsuperscript{$*$} & \textbf{0.6587}\textsuperscript{$*$} & \textbf{0.1534}\textsuperscript{$*$} & \textbf{0.3294}\textsuperscript{$*$} & \textbf{0.3905}\textsuperscript{$*$} & \textbf{0.3101}\textsuperscript{$*$} & \textbf{0.4176}\textsuperscript{$*$} \\
\midrule\midrule
\multirow{9}{*}{\centering{\textbf{Llama}}} & \multirow{7}{*}{\centering{non-fine-tuned}} & Q2D & 0.6082 & 0.6032 & 0.1459 & 0.2613 & 0.2833 & 0.2419 & 0.3573 \\
& & Q2E & 0.6385 & 0.5987 & 0.1460 & 0.3001 & 0.3554 & 0.2807 & 0.3866 \\
& & CoT & 0.5661 & 0.5847 & 0.1223 & 0.2552 & 0.2995 & 0.2168 & 0.3408 \\
& & GenQRE & 0.5093 & 0.4786 & 0.0978 & 0.2459 & 0.2195 & 0.1545 & 0.2843 \\
& & GenQRFusion & 0.6273 & 0.5474 & 0.1406 & 0.3109 & 0.3762 & 0.2889 & 0.3819 \\
& & GenCRF/SimDW\textsuperscript{$\dagger$} & 0.6517\textsuperscript{$*$} & \underline{0.6939}\textsuperscript{$*$} & 0.1505 & 0.3274\textsuperscript{$*$} & 0.3897\textsuperscript{$*$} & 0.3126 & 0.4210\textsuperscript{$*$} \\
& & GenCRF/ScoreDW\textsuperscript{$\dagger$} & 0.6561\textsuperscript{$*$} & 0.6824\textsuperscript{$*$} & 0.1539\textsuperscript{$*$} & 0.3286 & 0.3903\textsuperscript{$*$} & 0.3147\textsuperscript{$*$} & 0.4210\textsuperscript{$*$} \\
\cmidrule{2-10}
& \multirow{2}{*}{\centering{fine-tuned}} & GenCRF/ScoreDW-FT\textsuperscript{$\dagger$} & \underline{0.6566}\textsuperscript{$*$} & 0.6889 & \underline{0.1552}\textsuperscript{$*$} & \underline{0.3336}\textsuperscript{$*$} & \underline{0.3913} & \underline{0.3164}\textsuperscript{$*$} & \underline{0.4237}\textsuperscript{$*$} \\
& & GenCRF/ScoreDW-FT-QREM\textsuperscript{$\dagger$} & \textbf{0.6613}\textsuperscript{$*$} & \textbf{0.6952}\textsuperscript{$*$} & \textbf{0.1557}\textsuperscript{$*$} & \textbf{0.3357}\textsuperscript{$*$} & \textbf{0.3940}\textsuperscript{$*$} & \textbf{0.3199} & \textbf{0.4270}\textsuperscript{$*$} \\
\bottomrule[1pt]
\end{tabular}%
}}
\caption{nDCG@10 scores for GenCRF compared with multiple baselines across six datasets from the BEIR benchmark. Bold text for the best performance, underlined text for the second best. * denotes significant improvements (paired t-test with Holm-Bonferroni correction, p < 0.05) over the indicated baseline model(s). † denotes our proposed methods.}\label{tab:performance_comparison}
\end{table*}

\section{EXPERIMENTS}

\subsection{Setup}

We detail the experimental configuration, including datasets, baseline methods, and model specifications. We also detail the prompts used and specific parameters for each component of our framework.

\subsubsection{Experimental Datasets}
We conduct our main experiments on six datasets from the BEIR benchmark \cite{thakur2021beir} to evaluate retrieval performance: \emph{SciFact}, \emph{TREC-COVID}, \emph{SciDOCS}, \emph{NFCorpus}, \emph{DBPedia-entity}, and \emph{FiQA-2018}. For ablation studies and parameter analysis, we used two additional datasets: \emph{ArguAna} and \emph{CQADupStack-English}. The \emph{Quora} dataset is utilized for both constructing scoring data in the Fine-Tuning process and training our Query Evaluation Rewarding Model (QERM). 

\subsubsection{Models Used}
\paragraph{LLMs:} We employ Mistral-$7$B-Instruct-v$0.3$ and Llama-$3.1$-$8$B-Instruct models \cite{jiang2023, touvron2023}, with temperature $0.8$ and top\_p $0.95$ for diverse outputs. GPT-$4$o \cite{openai2024chatgpt} is used to generate high-quality reference scores for fine-tuning. We apply full-parameter fine-tuning to both models, using a learning rate of $1e-5$ for $5$ epochs, with a batch size of $16$.

\paragraph{Similarity Model:} SentenceBERT (all-mpnet-base-v2) \cite{reimers2019} is used to generate embeddings of initial query and generated queries. These embeddings are then used to calculate the cosine similarity between them within the GenCRF framework. We set a similarity threshold $\theta = 0.2$ to filter out irrelevant queries.\footnote{Similarity Threshold analysis in Appendix C.1.}

\paragraph{Retrieval Model:} MSMARCO-DistilBERT-base-TAS-B model is used for our retrieval step, which is specifically designed for Dense Passage Retrieval and trained on the MSMARCO passage dataset \cite{campos2016}, featuring 6-layer DistilBERT architecture optimized for retrieval.

\paragraph{Query Evaluation Rewarding Model:}
We use RoBERTa-Large Model \cite{liu2019} as QERM's backbone for its robustness in NLP tasks. The training uses a learning rate of $1e-5$ for 4 epochs, with a maximum of 2 iterations. The output logit threshold ($\varepsilon$) is set to the mean of the first iteration logits, ensuring an adaptive and contextually relevant baseline for query quality evaluation.

\subsubsection{Baseline Methods}
We compare our method against several established competitive baselines. For non-fusion methods, queries are structured as "initial query [SEP] generated query", where [SEP] is a separator token. The baseline methods include: Query2Doc (\emph{\textbf{Q2D}}): Generate pseudo-documents for query expansion \cite{wang2023c}; Query2Expansion (\emph{\textbf{Q2E}}): Expand queries with relevant keywords \cite{jagerman2023};
Query2CoT (\emph{\textbf{Q2C}}): Apply Chain of Thoughts for query reformulation \cite{wei2022}; GenQREnsemble (\emph{\textbf{GenQRE}}): Use multiple prompts to generate and concatenate keywords with initial query\cite{dhole2024a}; \textbf{\emph{GenQRFusion}}: Extend GenQREnsemble with keyword fusion method.

\subsubsection{Prompts Used}
Baseline methods utilize varying numbers of prompts: Q2D, Q2E, and Q2C each use four few-shot prompts, GenQRE uses ten \cite{dhole2024a}, and GenQR-Fusion randomly selects three prompts with a fusion strategy \cite{dhole2024b}. Our framework utilizes five types of prompts: three for diverse query generation (\emph{Contextual Expansion}, \emph{Detail Specific}, \emph{Aspect Specific}), one for \emph{Clustering-Generation}\footnote{Cluster analysis in Appendix D.1. and Appendix D.2.}, and one for \emph{Scoring}. The Scoring prompt evaluates generated queries based on Relevance, Specificity, Clarity, Comprehensiveness, and Usefulness, assigning scores from $1$ to $100$, with a threshold $\theta = 60$ to ensures only high-scoring reformulations contribute to the final aggregated query.\footnote{Score Threshold analysis in Appendix C.2.} Detailed descriptions of all prompts are provided in the appendix.

\subsection{Experimental Analysis}

In our experiments, we evaluated the performance of our proposed GenCRF framework across six datasets from the BEIR benchmark, as shown in Table \ref{tab:performance_comparison}. Ensemble-based methods such as GenQRE and GenQRFusion outperforms single prompted methods on average, with GenQRFusion demonstrating particularly strong results. This indicates that ensemble based apporaches using multiple prompts to expand retrieval queries enhance information gaining and improve retrieval performance.

However, our proposed GenCRF methods, such as SimDW and ScoreDW, further improve upon these ensemble-based approachs. Our strategies consistently outperform GenQRE and GenQRFusion across all datasets. This result demonstrates the effectiveness of both our multi-intent query generation and dynamic weight aggregation techniques, offering an effective approach compared to static weighting strategies in advance. We provide a more detailed analysis and examination of these two components in Section 4.3.1. and Section 5.

Additionally, the fine-tuned method ScoreDW-FT demonstrates stronger performance across all datasets, indicating fine-tuning process enhances the LLM's consistent and scalable quality assessment. Moreover, ScoreDW-FT-QERM consistently achieves the best results among all methods. It effectively guides the LLMs in the query refinement process by iteratively assessing GenCRF based on nDCG@10 scores, thereby enhancing the overall adaptability of our GenCRF framework. The improvements are most pronounced in trec-covid-beir and dbpedia-entity, highlighting the robustness of our approach across various retrieval tasks.

\subsection{Ablation Studies}
To validate GenCRF's robustness, we conduct ablation studies on key parameters using \emph{ArguAna} and \emph{CQADupStack-English} datasets. For comparison with other methods, particularly those that do not use weighted aggregation, we introduce Direct Concatenation (DC) method:
\begin{equation}
q_{\text{agg}}^{\text{DC}} = q_{\text{init}} + \text{[SEP]} + \sum_{i=1}^{|\mathcal{Q}{\text{final}}|} (q_{\text{final}, i} + \text{[SEP]})
\end{equation}

DC combines the initial query $q_{\text{init}}$ with all reformulated queries using [SEP] tokens as separators. Also, we introduce Fixed Weights (FW) method for ablation study:
\begin{equation}
q_{\text{agg}}^{\text{FW}} = w_0 \cdot q_{\text{init}} + \frac{1-w_0}{|\mathcal{Q}{\text{final}}|} \sum_{i=1}^{|\mathcal{Q}{\text{final}}|} q_{\text{f},i}
\end{equation}

Here, $w_0$ is the fixed weight for the initial query, and $(1-w0)/|\mathcal{Q}_{\text{final}}|$ is the equal weight applied to each reformulation query.

\paragraph{4.3.1. Initial Query Weight.} 

\begin{figure}
    \vspace{-0.8em}
    \centering
    \includegraphics[width=1\linewidth]{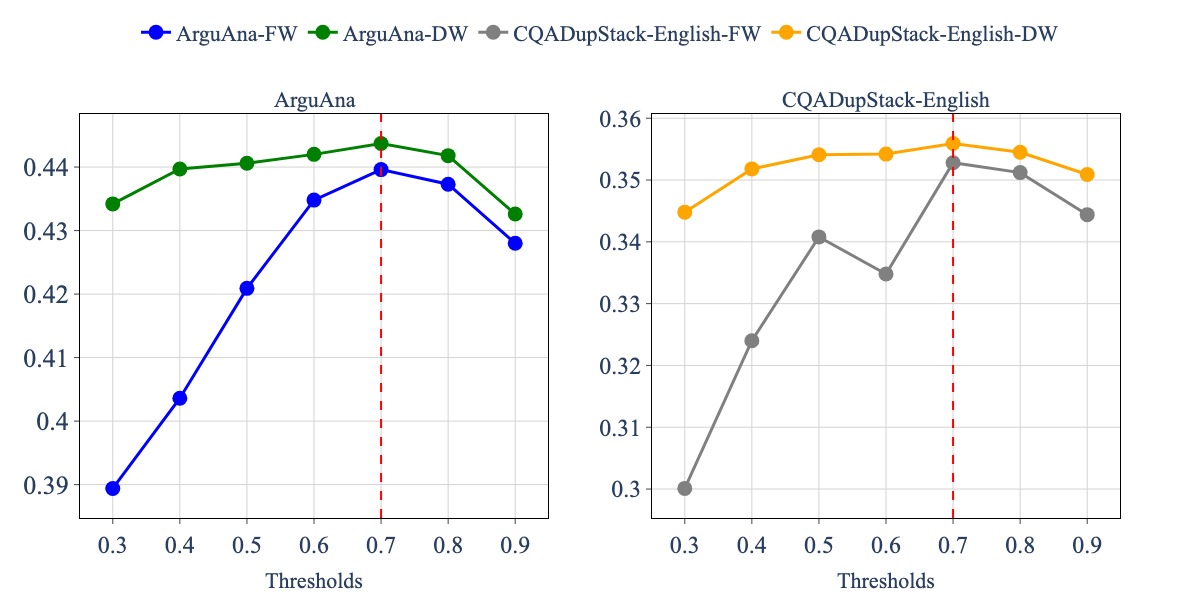}
    \caption{Initial Weight Comparison of FW and DW}
    \label{fig:initial_weight_comparison}
    \vspace{-0.5em}
\end{figure}

To determine the optimal weight for Weighted Aggregation, we investigate on both Fixed Weights (FW) and ScoreDW/FT strategies. We vary $w_0$ from $0.3$ to $0.9$ in 0.1 increments, evaluating nDCG@10 for both strategies. As shown in Figure \ref{fig:initial_weight_comparison}, $w_0=0.7$ achieves optimal performance across both strategies and datasets. For fairness in building baseline, we applied this optimal $w_0 = 0.7$ to GenQRFusion method as well. 

\begin{figure}
    \centering
    \includegraphics[width=1\linewidth]{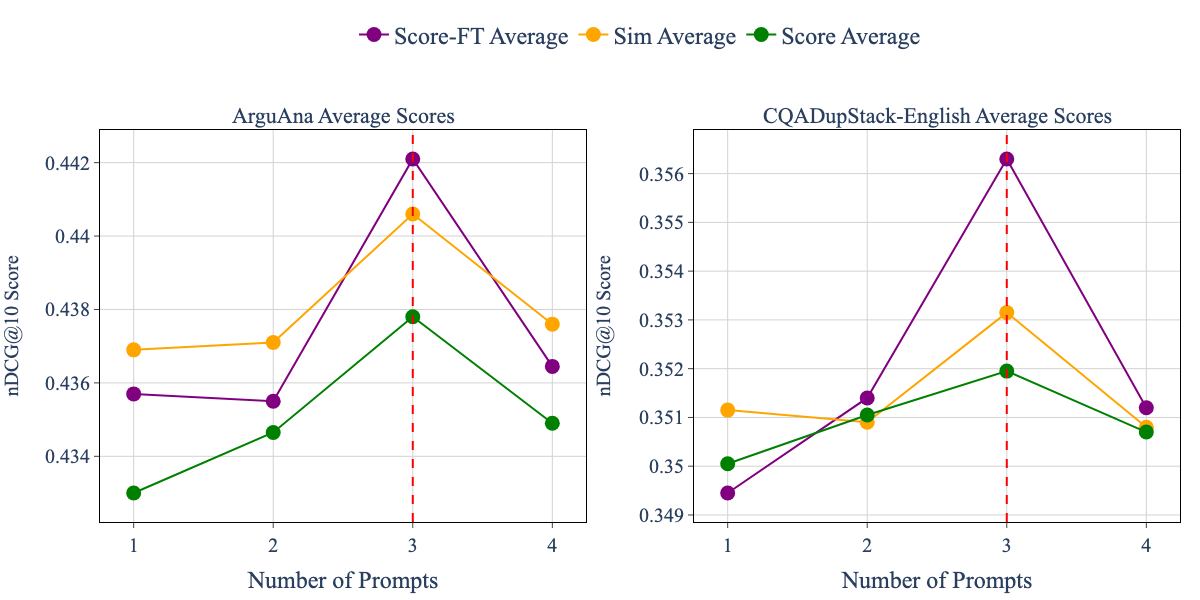}
    \caption{nDCG@10 scores for different prompt quantities}
    \label{fig:prompt_quantity}
\end{figure}

\paragraph{4.3.2. Impact of Prompt Quantity.} 
We investigate the effect of varying the number of prompt types on retrieval performance, measured by nDCG@10 scores. The prompts included "contextual expansion," "detail specific," "aspect specific," and "clarity enhancement."\footnote{Clarity Prompt in Appendix B.3.} We evaluated all possible combinations of these prompts and calculated the average performance for each number of prompts used. As shown in Figure \ref{fig:prompt_quantity}, performance typically improves when increasing from $1$ to $3$ prompts, but adding a fourth prompt does not lead to further enhancements and conversely decreases performance. Thus, we selected $3$ prompt types as the optimal configuration for maximizing retrieval performance in this study.

\begin{table}[htbp]
    \centering
    \small
    \begin{tabular}{@{}c@{\hspace{0.5em}}c@{\hspace{0.5em}}c@{\hspace{0.5em}}c@{}}
    \toprule
    \textbf{Datasets} & \textbf{Gen(N)} & \textbf{SimDW} & \textbf{ScoreDW-FT} \\ 
    \midrule
    \multirow{4}{*}{\emph{\textbf{ArguAna}}} 
    & 1 & 0.4361 &  0.4367 \\ 
    & \textbf{2} &  \textbf{0.4395} & \textbf{0.4421} \\ 
    & 3 & \underline{0.4366} & \underline{0.4395} \\ 
    & 4 & 0.4351 & 0.4369  \\ 
    \midrule
    \multirow{4}{*}{\emph{\textbf{CQADupStack-English}}}
    & 1 & 0.3251 & 0.3478  \\ 
    & \textbf{2} & \textbf{0.3543} & \textbf{0.3563} \\ 
    & 3 &  \underline{0.3482} & \underline{0.3498} \\ 
    & 4 & 0.3491 & 0.3497 \\ 
    \bottomrule
    \end{tabular}
    \caption{Comparison of nDCG@10 scores for different numbers of generated queries \(N\) using SimDW and ScoreDW-FT strategies.}
    \label{tab:generated-results-2}
\end{table}

\paragraph{4.3.3. Impact of Generated Query Count.} We explore the effect of varying the number of generated queries \(N\) per prompt on retrieval performance, as described in Section 3.2. Experiment were conducted with \(N\) ranging from $1$ to $4$ for both SimDW and ScoreDW-FT strategies across \emph{ArguAna} and \emph{CQADupstack-English} dataasets. As shown in Table \ref{tab:generated-results-2}, generating $2$ queries per prompt consistently yields the best performance across both datasets and strategies. The performance decline for \(N\) suggests that additional generation may introduce noise or redundancy, which may be attributed to the excessive length of single-prompt generated responses or their mutual interference.

\paragraph{4.3.4. Iterative Optimization with QERM.} We examine the impact of iterative optimization using the Query Evaluation Rewarding Model (QERM) on our ScoreDW-FT-QERM framework, with iteration counts ranging from $1$ to $4$, as shown in table \ref{tab:generated-results}. We observe that the best iteration count is $2$, significantly surpassing the score where iteration count is $4$. The result indicate that the integration of QERM with two iterations achieve an optimal result, allowing the ScoreDW-FT-QERM framework to adaptively optimize query generation and clustering, resulting in more precise and relevant retrieval outcomes across diverse datasets.

\begin{table}[htbp]
    \centering
    \small
    \begin{tabular}{@{}c@{\hspace{0.5em}}c@{\hspace{0.5em}}c@{}}
    \toprule
    \textbf{Datasets} & \textbf{Iteration} & \textbf{nDCG@10} \\ 
    \midrule
    \multirow{4}{*}{\emph{\textbf{ArguAna}}} 
    & 1 & \underline{0.4397}  \\ 
    & \textbf{2} & \textbf{0.4421}  \\ 
    & 3 & 0.4369 \\ 
    & 4 & 0.4358  \\ 
    \midrule
    \multirow{4}{*}{\emph{\textbf{CQADupStack-English}}}
    & 1 & 0.3527 \\ 
    & \textbf{2} & \textbf{0.3563} \\ 
    & 3 & \underline{0.3540} \\ 
    & 4 & 0.3536 \\ 
    \bottomrule
    \end{tabular}
    \caption{nDCG@10 scores for different QERM iteration counts using ScoreDW-FT-QERM.}
    \label{tab:generated-results}
\end{table}
\vspace{-1.5em}

\begin{figure}
    \centering
    \includegraphics[width=1\linewidth]{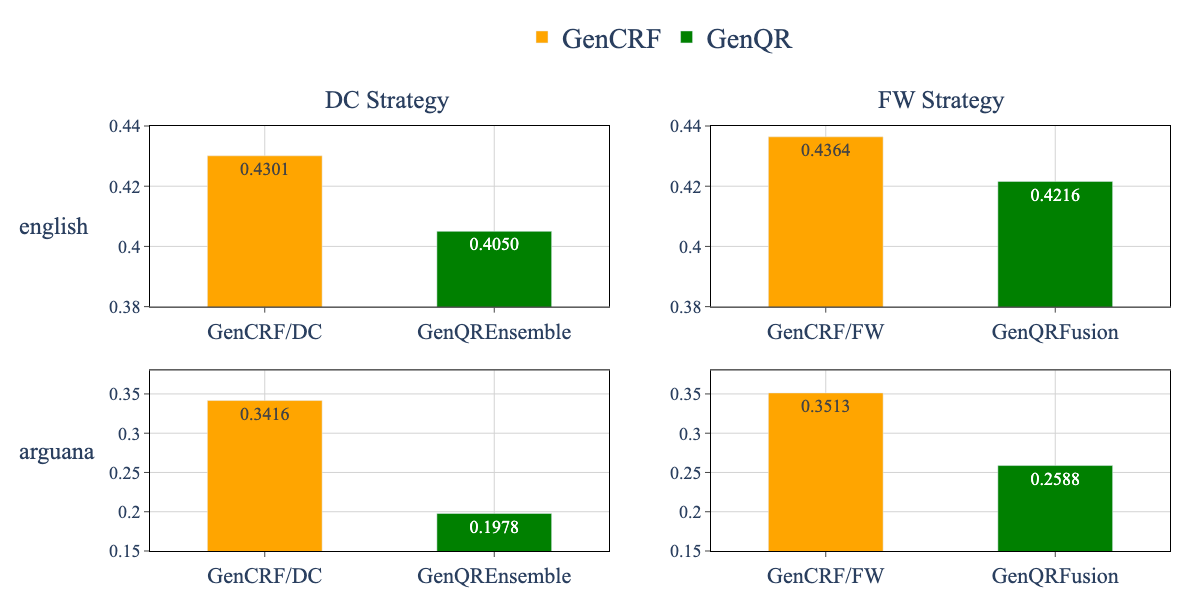}
    \caption{Performance comparison of GenCRF with GenQR using DC and FW strategies}
    \label{fig:enter-label1}
\end{figure}

\begin{figure}
    \centering
    \includegraphics[width=1\linewidth]{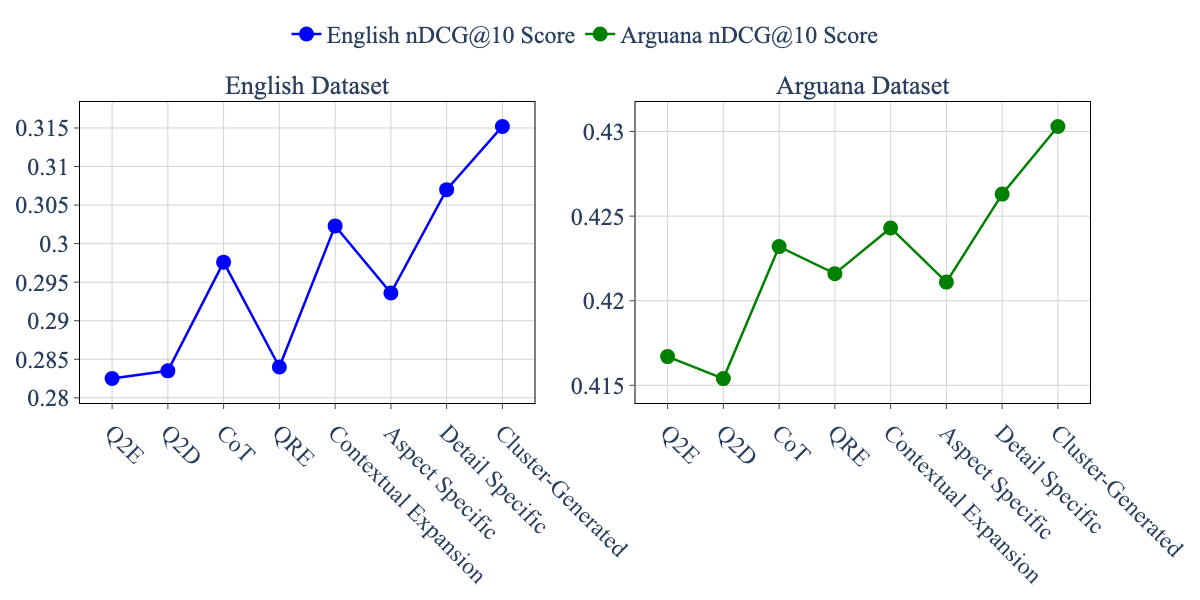}
    \caption{Comparsion of nDCG@10 scores between GenCRF's prompts and baseline methods}
    \label{fig:enter-label2}
\end{figure}

\section{Generation study and Discussions}
Our GenCRF has demonstrated superior performance in baseline comparisons, effectively capturing query intent compared to existing methods. Figure \ref{fig:enter-label1} reveals 
GenCRF outperforming GenQR methods, even with basic aggregation strategies such as Direct Concatenation (DC) and Fixed Weights (FW). While GenQRFusion relies on keyword-based methods that often fails to capture the underlying query intent, GenCRF's prompts explore various query facets, resulting in more comprehensive reformulations that capture nuances keyword-based methods neglect.
 
As shown in Figure \ref{fig:enter-label2}, our individual prompts (\emph{Contextual Expansion, Detail Specific and Aspect Specific}) outperform baseline methods such as Q2E, Q2D and CoT. Our prompts capture deeper query semantics, contrasting with conventional methods' focus on surface-level information. Notably, our Cluster-Generated method, which combines diverse insights from various prompts, achieves the best results, demonstrating the effectiveness of integrating multiple perspectives in query reformulation-an approach absent in single-prompt methods.

\section{Conclusion}
We present the Generative Clustering and Reformulation Framework (GenCRF), which demonstrates significant advancements over existing competitive baseline methods, achieving up to 12\% increase on BEIR benchmark. Our approach combines diverse prompting strategies and clustering refinement to accurately capture and reformulate query intents. We introduced our optimization techniques including weighted aggregation methods: \emph{SimDW}, \emph{ScoreDW}, \emph{ScoreDW-FT} and the evaluation rewarding model \emph{QERM}, enhancing GenCRF's performance and offering a more precise, user-centric information retrieval experience. Extensive ablation studies have confirmed the reasonableness and robustness of the GenCRF framework by exploring key parameters and settings across datasets from the BEIR benchmark. Future work could explore GenCRF's application to real-world search scenarios, potentially enhancing its effectiveness in practical information retrieval contexts.

\bibliography{custom}
\clearpage
\appendix
\section{Appendix A. Overview}

This appendix provides a comprehensive overview of the methodologies and experimental setups employed in our study. We detail the prompts used in our baseline models and our GenCRF framework, including those used for ablation studies. Additionally, we present our methods for finding optimal simlilarity and score thresholds, also conduct a cluster anaylsis within the GenCRF framework.

\section{Appendix B. Prompts}
We share five prompts utilized in our experiments, including: Query2Doc (\emph{\textbf{Q2D}}): Generate pseudo-documents and expands queries \cite{wang2023c}; Query2Expansion (\emph{\textbf{Q2E}}): Expand queries with relevant keywords \cite{jagerman2023}; Query2CoT (\emph{\textbf{Q2C}}): Reformulate queries based on Chain of Thoughts prompting \cite{wei2022}; GenQREnsemble (\emph{\textbf{GenQRE}}): Applies multiple prompts to generate various keyword sets concatenated within the initial query\cite{dhole2024a} and GenCRF (Ours).

\subsection{Q2D, Q2E, Q2C}

\begin{table}[H]
    \vspace{-0.2em}
    \centering
    \small{  
    \scalebox{0.8}{ 
    \begin{tabular}{p{8cm}} \hline
        \textbf{Query2Doc} \\ \hline
        Write a passage that answers the given query:\\ \\
        Query: \{query 1\} \\
        Passage: \{doc 1\} \\ \\

        Query: \{query 2\} \\
        Passage: \{doc 2\} \\ \\

        Query: \{query 3\} \\
        Passage: \{doc 3\} \\ \\

        Query: \{query 4\} \\
        Passage: \{doc 4\} \\ \\

        Query: \{query\} \\
        Passage: \\ \hline
    \end{tabular}
    }
    }
    \caption{Prompt for Q2D}
    \label{tab:contextual-expansion1}
    \vspace{-0.2em}
\end{table}

\begin{table}[H]
    \vspace{-0.2em}
    \centering
    \small{  
    \scalebox{0.8}{  
    \begin{tabular}{p{8cm}} \hline
        \textbf{Query2Expansion} \\ \hline
        Write a list of keywords for the given query: \\ \\
        Query: \{query 1\} \\
        Keywords: \{expansion 1\} \\ \\

        Query: \{query 2\} \\
        Keywords: \{expansion 2\} \\ \\

        Query: \{query 3\} \\
        Keywords: \{expansion 3\} \\ \\

        Query: \{query 4\} \\
        Keywords: \{expansion 4\} \\ \\

        Query: \{query\} \\
        Keywords: \\ \hline
    \end{tabular}
    }
    }
    \caption{Prompt for Q2E}
    \label{tab:contextual-expansion2}
    \vspace{-0.2em}
\end{table}

\begin{table}[H]
    \vspace{-0.2em}
    \centering
    \small{  
    \scalebox{0.9}{  
    \begin{tabular}{p{8cm}} \hline
        \textbf{Query2CoT} \\ \hline
        Let's think step by step. \\
        Answer the following query, and give the rationale before answering. Below is the query: \\
        \{query\} \\ \hline
    \end{tabular}}
    }
    \caption{Prompt for Q2C}
    \label{tab:contextual-expansion3}
    \vspace{-0.2em}
\end{table}

\subsection{GenQRE}
\begin{table}[H]
    \vspace{-0.2em}
    \centering
    \small{ 
    \scalebox{0.9}{ 
    \begin{tabular}{p{8cm}} \hline
        \textbf{GenQREnsemble} \\ \hline
        1 Improve the search effectiveness by suggesting expansion terms for the query. \\
        2 Recommend expansion terms for the query to improve search results. \\
        3 Improve the search effectiveness by suggesting useful expansion terms for the query. \\
        4 Maximize search utility by suggesting relevant expansion phrases for the query. \\
        5 Enhance search efficiency by proposing valuable terms to expand the query. \\
        6 Elevate search performance by recommending relevant expansion phrases for the query. \\
        7 Boost the search accuracy by providing helpful expansion terms to enrich the query. \\ 
        8 Increase the search efficacy by offering beneficial expansion keywords for the query. \\
        9 Optimize search results by suggesting meaningful expansion terms to enhance the query. \\ 
        10 Enhance search outcomes by recommending beneficial expansion terms to supplement the query.
        \\ \hline
    \end{tabular}}
    }
    \caption{Prompt for GenQRE}
    \label{tab:contextual-expansion4}
    \vspace{-0.2em}
\end{table}

\subsection{GenCRF}
\begin{table}[H]
    \vspace{-0.2em}
    \centering
    \small{ 
    \scalebox{0.9}{ 
    \begin{tabular}{p{8cm}} \hline
        \textbf{Contextual Expansion} \\ \hline
        You are a contextual expansion expert. Your task is to understand the core intent of the original query and provide a refined, contextually expanded answer. Provide a clear and concise response based on the original query. \\
        Below is the query: \{...\} \\ \hline
    \end{tabular}}
    }
    \caption{Prompt for Contextual Expansion}
    \label{tab:contextual-expansion5}
    \vspace{-0.2em}
\end{table}

\begin{table}[H]
    \vspace{-0.2em}
    \centering
    \small{ 
    \scalebox{0.9}{ 
    \begin{tabular}{p{8cm}} \hline
        \textbf{Detail Specific} \\ \hline
        You are a detail-specific expert. Your task is to understand the core intent of the original query and provide a refined, detailed answer focusing on particular details or subtopics directly related to the query. Provide a clear and concise response based on the original query.\\ 
        Below is the query: \{...\} \\ \hline
    \end{tabular}}
    }
    \caption{Prompt for Detail Specific}
    \label{tab:contextual-expansion6}
    \vspace{-0.2em}
\end{table}

\begin{table}[H]
    \vspace{-0.2em}
    \centering
    \small{  
    \scalebox{0.9}{  
    \begin{tabular}{p{8cm}} \hline
        \textbf{Aspect Specific} \\ \hline
        You are an aspect-specific inquiry expert. Your task is to understand the core intent of the original query and provide a refined answer focusing on a specific aspect or dimension within the topic. Provide a clear and concise response based on the original query. \\
        Below is the query: \{...\} \\ \hline
    \end{tabular}}
    }
    \caption{Prompt for Aspect Specific}
    \label{tab:contextual-expansion7}
    \vspace{-0.2em}
\end{table}

\begin{table}[H]
    \vspace{-0.2em}
    \centering
    \small{  
    \scalebox{0.9}{  
    \begin{tabular}{p{8cm}} \hline
        \textbf{Clarity Enhancement} \\ \hline
        You are a clarity-enhancement expert. Your task is to understand the core intent of the original query and reformulate it to enhance clarity and specificity. Focus on eliminating ambiguity and ensuring the query is straightforward, which aids in retrieving the most relevant contexts. Provide a clear and concise response based on the original query. \\
        Below is the query: \{...\} \\ \hline
    \end{tabular}}
    }
    \caption{Prompt for Clarity Enhancement}
    \label{tab:clarity_enhancement}
    \vspace{-0.2em}
\end{table}

\begin{table}[H]
    \vspace{-0.2em}
    \centering
    \small{ 
    \scalebox{0.9}{ 
    \begin{tabular}{p{8cm}} \hline
        \textbf{Clustering Refinement} \\ \hline
        You are an expert in clustering and query refinement. Your task is to review the original query alongside the generated queries, and then cluster them into 1 to 3 groups based on their similarity and relevance. \\
        The number of clusters should be determined dynamically. Focus primarily on the relationship of the generated queries to the original query. For each identified cluster, provide only one refined query that incorporates elements from the original and generated queries within that cluster with useful information for document retrieval. \\
        The output should be presented in JSON format, structured as follows:
        \{'cluster1': 'refined\_query\_1', 'cluster2': 'refined\_query\_2', 'cluster3': 'refined\_query\_3'\} \\
        The output must be restricted to 1 to 3 groups. \\
        Below is the query: \{...\} \\ \hline
    \end{tabular}}
    }
    \caption{Prompt for Clustering Refinement}
    \label{tab:contextual-expansion8}
    \vspace{-0.2em}
\end{table}

\begin{table}[H]
    \vspace{-0.2em}
    \centering
    \small{ 
    \scalebox{0.9}{ 
    \begin{tabular}{p{8cm}} \hline
        \textbf{LLM Scoring} \\ \hline
        You are an expert in scoring cluster queries. Evaluate the clustering of queries using the following criteria for each cluster: Relevance, Specificity, Clarity, Comprehensiveness, and Usefulness for retrieval.\\
        Assign a score from 1 to 100, where 1 is the lowest and 100 is the highest performance in relation to the original query. Avoid defaulting to high scores unless they are clearly justified. Carefully consider both the strengths and weaknesses of each cluster. \\
        For instance, a cluster with relevant but not highly specific results might score between 40 and 60, while a cluster that is both highly relevant and specific might score between 70 and 100. Conversely, a cluster lacking clarity or comprehensiveness should score lower, between 10 and 30. \\
        Provide scores that accurately reflect the variation in quality across clusters. List your scores for each cluster in the following format: [score\_cluster1, score\_cluster2, score\_cluster3]. \\
        Return your scores in a list format only, without additional commentary. \\ 
        Initial Query: \{...\} \\
        Cluster-Generated Queries : \{...\} \\ \hline
    \end{tabular}}
    }
    \caption{Prompt for Scoring}
    \label{tab:contextual-expansion9}
    \vspace{-0.2em}
\end{table}

\section{Appendix C. Similarity and Score Threshold for Dynamic Weights}
We conducted comprehensive experiments to determine the optimal similarity and score thresholds. 
\subsection{Similarity Threshold}
\begin{figure}
    \centering
    \includegraphics[width=1\linewidth]{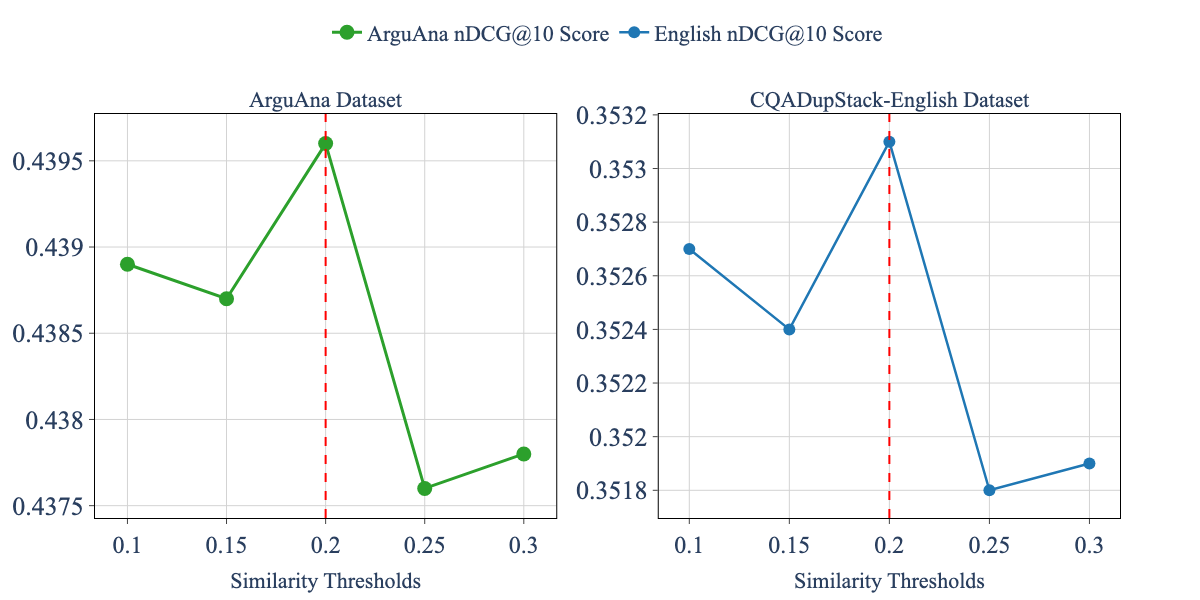}
    \caption{Impact of similarity thresholds on nDCG@10 scores}
    \label{fig:sim_threshold}
\end{figure}

We experimented with similarity ranging from 0.1 to 0.3 to identify the optimal threshold. As illustrated in Figure \ref{fig:sim_threshold}, a threshold of 0.2  yielded the highest nDCG@10 score across all the ablation datasets. This finding demonstrates that an appropriate threshold can effectively enhance the retrieval performance and that the threshold is generalizable across different testsets.

\subsection{Score Threshold}
\begin{figure}
    \centering
    \includegraphics[width=1\linewidth]{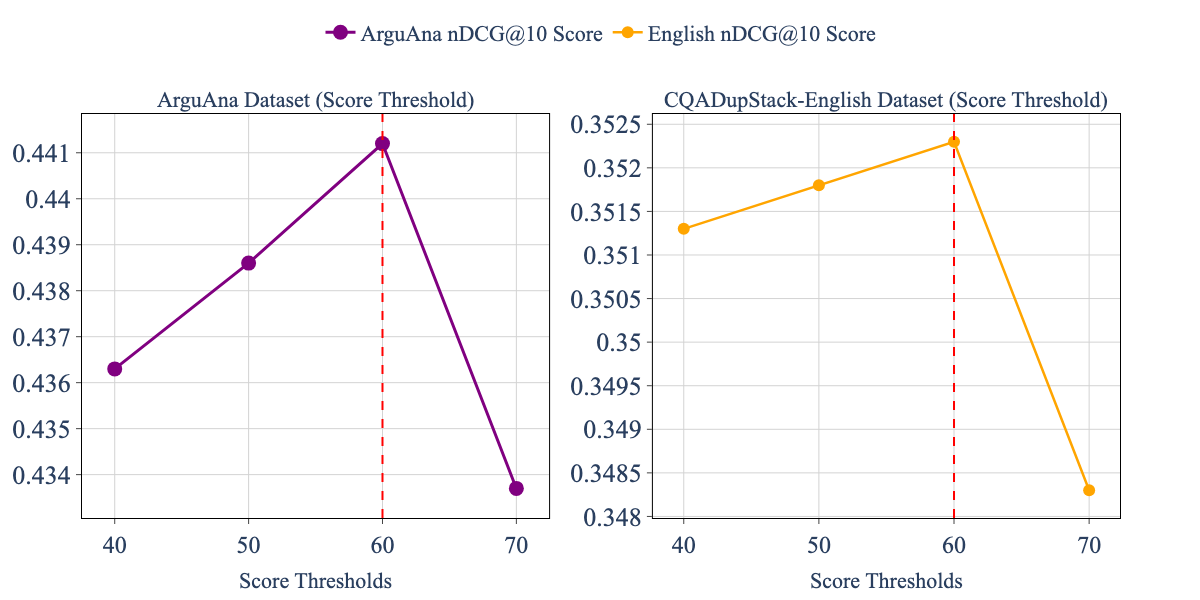}
    \caption{Impact of score thresholds on nDCG@10 scores}
    \label{fig:score_threshold}
\end{figure}

We also investigated the impact of various score thresholds on the performance of our model. Figure \ref{fig:score_threshold} shows the results of experiments with score thresholds ranging from 40 to 70. For both datasets, a score threshold of 60 resulted in the highest nDCG@10 scores, which suggests that our threshold effectively filters our lower-quality generated queries while retaining those that contribute most to improved retrieval performance.

\section{Appendix D. Cluster Analysis}
We analyze how the GenCRF framework clusters data across our main datasets, focusing on the distribution of cluster counts and the similarity between clusters. 

\subsection{Cluster Counts}
\begin{figure}[!htpb]
    \centering
    \includegraphics[width=0.9\linewidth]{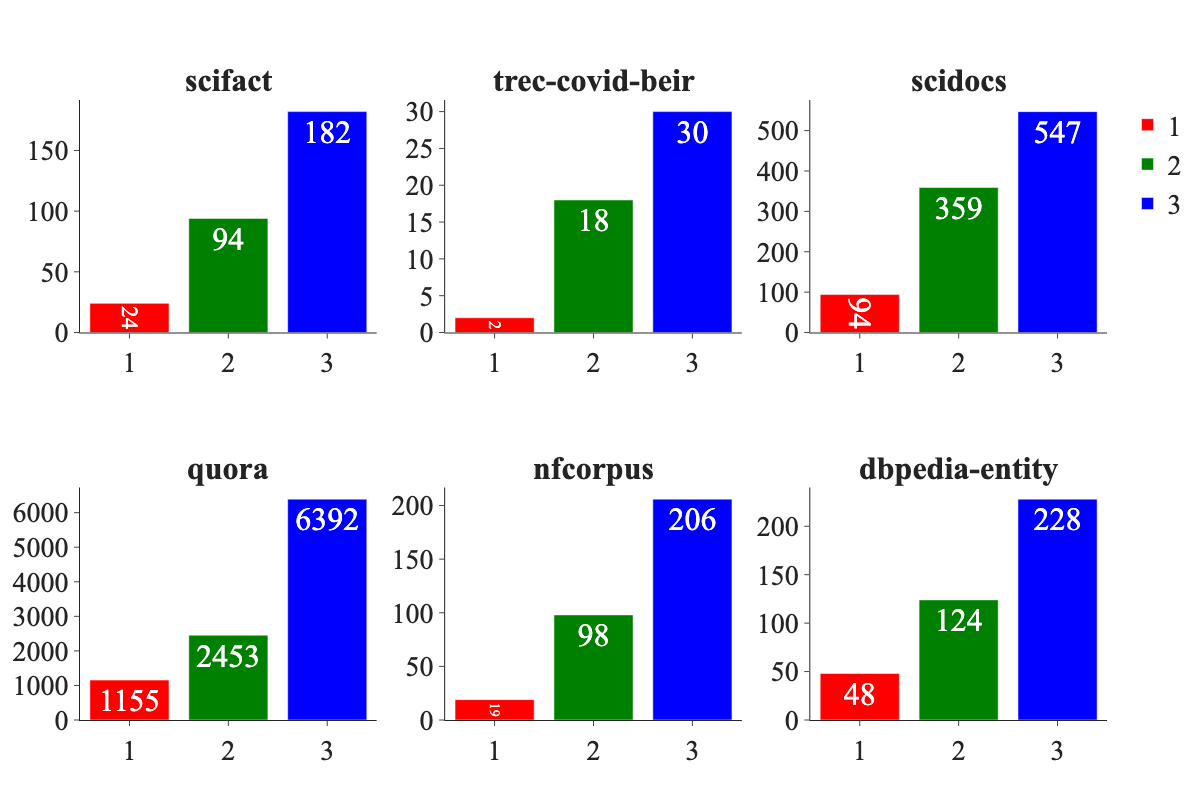}
    \caption{Distribution of Cluster Counts Across Datasets}
    \label{fig:cluster_distribution}
\end{figure}

As shown in Figure \ref{fig:cluster_distribution}, the GenCRF framework predominantly form three clusters across all datasets, followed by two clusters, with a small proportion of single clusters. The result indicates that our framework often identifies multiple distinct aspects of query intents. This multi-faceted clustering approach likely contributes to the framework's ability to generate diverse and comprehensive query reformulations.

\subsection{Similarity between Clusters}
\begin{figure}[!htpb]
    \centering
    \includegraphics[width=0.9\linewidth]{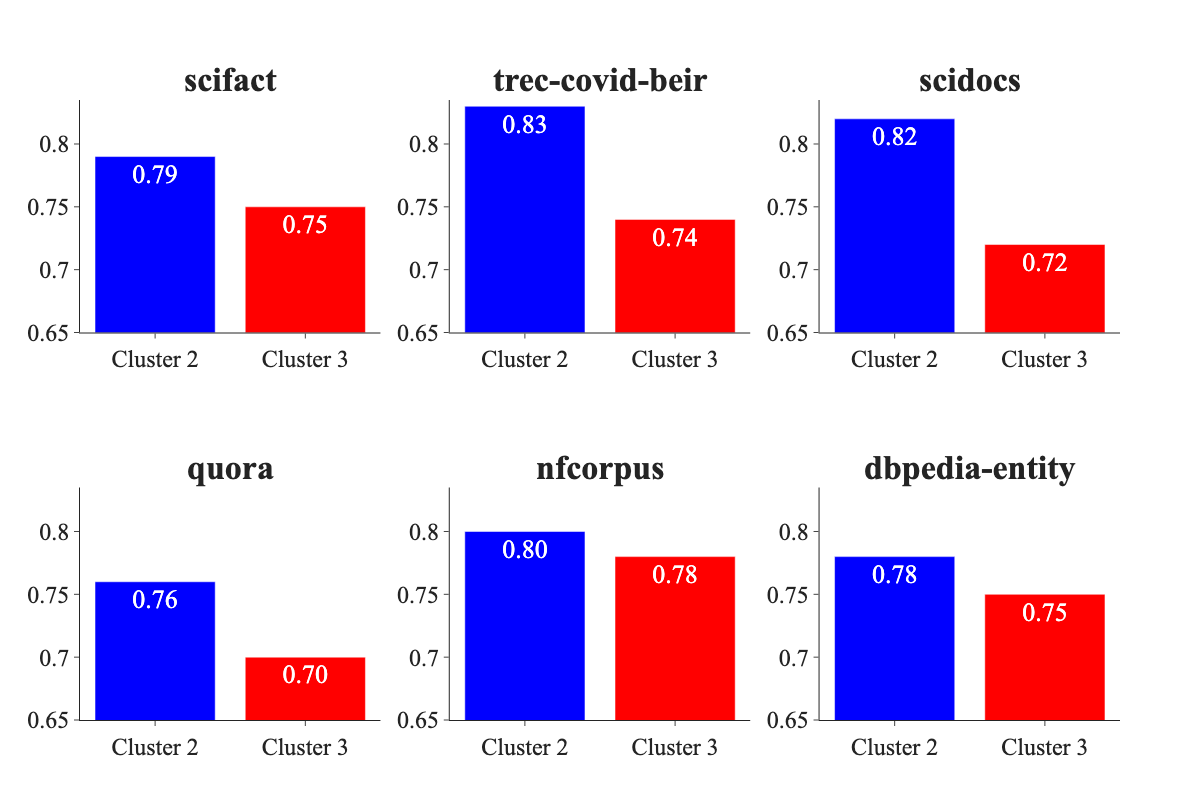}
    \caption{Similarity between Clusters}
    \label{fig:cluster_similarity}
\end{figure}

Figure \ref{fig:cluster_similarity} illustrates the similarity between clusters when two or three clusters are formed. We observe that $2$ cluster formation consistently show higher similarity scores compared to three cluster formations, and the similarity scores for both $2$ cluster and $3$ cluster formations are relatively high, indicating greater diversity, and making these clusters more effective at capturing different aspects of query intents.

\end{document}